\documentclass[12pt,psamsfonts]{amsart}
\usepackage{euscript,amsmath,amsthm,amsfonts,amssymb}
\usepackage{tikz}
\usepackage{graphicx}
\usepackage[all,knot]{xy}
\xyoption{arc}
\usepackage{epsfig}
\usepackage[dvips]{pstricks}

\newcommand{\p}{\textsf{p}}

\newcommand{\Q}{\mathbb Q}
\newcommand{\R}{\mathbb{R}}
\newcommand{\Z}{\mathbb{Z}}

\newcommand{\one}{\mathbf{1}}
\newcommand{\C}{\mathbb C}
\newcommand{\mC}{\mathcal{C}}

\newcommand{\mZ}{\mathcal{Z}}

\numberwithin{equation}{section}

\theoremstyle{definition}

\begin{document}
\title[TQFT and TI]
{(3+1)-TQFTs and Topological Insulators}

\author{Kevin Walker and Zhenghan Wang}
\email{kevin@canyon23.net\\zhenghwa@microsoft.com}
\address{Microsoft Station Q\\CNSI Bldg Rm 2243\\
    University of California\\
    Santa Barbara, CA 93106-6105\\
    U.S.A.}

\thanks{The second author is grateful to Xiao-Gang Wen for very insightful comments on an earlier draft of the paper.  We thank L. Chang for drawing most of the pictures.}

\begin{abstract}

Levin-Wen models are microscopic spin models for topological phases of matter in $(2+1)$-dimension.
We introduce a generalization of such models to $(3+1)$-dimension based on unitary braided fusion categories, also known as unitary
premodular categories.  We discuss the ground state degeneracy on $3$-manifolds and statistics of excitations which include both points and
defect loops.  Potential connections with recently proposed fractional topological insulators and projective ribbon permutation statistics are described.


Pacs: 02.40.Pc, 03.65.Vf, 03.75.Lm

\end{abstract}
\maketitle

\section{Introduction}

In a remarkable paper, M.\ Levin and X.-G.\ Wen defined a family of rigorously solvable lattice spin Hamiltonians for a large class of $2D$ topological phases of matter based on string-net, called the Levin-Wen model \cite{LW}.  The Levin-Wen model takes a unitary fusion category $\mC$ as input, and as output, realizes the topological phase effectively described by the topological quantum field theory (TQFT) based on the Drinfeld center $Z(\mC)$ of $\mC$.  In Section $V$ of their paper, Levin-Wen model is generalized to $3D$ (and $\geq 4$ dimensions) using unitary symmetric fusion categories.  The $3D$ Levin-Wen model realizes all discrete gauge theories coupled with bosons and fermions.
In this paper, we show that a generalized string-net model exists for all unitary braided fusion categories, which include the unitary symmetric fusion categories and unitary modular categories as two special cases.  The case for unitary modular categories is a generalization of the $(3+1)$-BF theories.  A unitary braided fusion category, also called a unitary premodular category, is some non-trivial product of a discrete gauge theory with a unitary modular category.  Since unitary modular categories are algebraic theories of anyons, therefore our new models can be thought as discrete gauge theories coupled with anyons.

As a generalization of the Levin-Wen model, our Hamiltonians also consist of two kinds of commuting projectors, and are stable under small, yet  arbitrary perturbations.  On the $3$-sphere, the ground state manifold is non-degenerate.  In general the discrete gauge theory part of a unitary premodular category corresponds to a finite abelian group $G$, and we expect the ground state manifold of our Hamiltonian on a general $3$-manifold $X$ is isomorphic to $\mathbb{C}[H_1(X;G)]$.  The ground state manifold on any $3$-manifold $X$ are examples of $3D$-error correction codes.  It will be interesting to know if there are examples of self-correcting quantum memories for some theories on certain $3$-manifolds.  Pointed excitations in our models are still bosons and fermions.  What interesting is the existence of extended excitations such as loops and $\theta$-graphs.  The mutual statistics of pointed excitations and the extended excitations are more general than bosonic and fermionic statistics.

When the time reversal symmetry of fractional topological insulators is broken, fractional topological insulators can be connected to some topological phases including the trivial one.  Therefore, $3D$ fractional topological insulators can be considered as topological phases with symmetry \cite{Wen02}.  We conjecture that our $(3+1)$-TQFTs are the underlying topological orders for $3D$ topological insulators and their generalizations.  A classification of the compatible symmetries in our models using the projective symmetry group \cite{Wen03} should reveal a connection to topological orders with symmetry.  A classification of time-reversal symmetries within a given unitary premodular category would lead to a classification of all $3D$-topological insulators with the same underlying topological order.

\section{Topology in Condensed Matter Physics}

As a branch of pure mathematics, topology is the study of spaces regardless of metric.  Therefore,
topology in physics seems to be a strange occurrence because distance and time are of paramount importance in
physical measurements.  But the first application of topology in physics predates topology and goes back at least to $1833$ when Gauss revealed his
beautiful formula for the linking number.  It was argued that the formula originated first from Gauss's study of the tracking of asteroids and comets, and later he applied his formula to electromagnetism \cite{Epple}.

Suppose $L_1, L_2$ are  two disjoint simple closed curves in $\R^3$.  If
there is an electric current of strength $j_1$ in the wire $L_1=\{x'(s)|s\in S^1\}$, then it generates
a magnetic field
$$B(x)=\frac{\mu_0 j_1}{4\pi} \oint_{L_1} \frac{(x'-x)\times dx'}{|x'-x|},$$
by the Biot-Savant law for $x\in \R^3\backslash L_1$.  By the Maxwell equation $\oint_{L_2} B(x) dx=\mu_0 j$ for $L_2=\{x(t)|t\in S^1\}$, it follows that
$$\mu_0 j_2= \frac{\mu_0 j_1}{4\pi} \oint_{L_2}\oint_{L_1} \frac{(x'-x)\cdot (dx'\times dx)}{|x'-x|}.$$
Therefore, an electromagnetic definition of the linking number between $L_1$ and $L_2$ will be the ratio $\frac{j_2}{j_1}$.
Gauss's formula has a geometric explanation as follows.  Taking two points $x',x$ on $L_1, L_2$ and normalizing the line segment $x'$-$x$ to $\frac{x'-x}{|x'-x|}$, then we obtain a map from the abstract torus $S^1\times S^1$ of $s,t$ parameters to the unit sphere $S^2\subset \R^3$.  Gauss's formula is simply the degree of this map.  Though linking number is a very useful invariant for linkage, Maxwell, who were very interested in knot theory,  seemed to be the first to find two un-separable disjoint simple closed curves with linking number equal to $0$ \cite{Epple}.

An important topic in topology is the study of special spaces called manifolds.  An $n$-manifold $M$ is a space that is locally
the Euclidean space $\R^n$ up to homeomorphism.
Simple examples are $n$-spheres: $1$-sphere is the circle,  and $2$-sphere is our ordinary sphere.
It is a challenge to visualize manifolds beyond dimension $2$, and see their relevance to condensed matter physics.
After all, all physical experiments are carried out in our $3$-space $\R^3$, and most $3$-manifolds cannot be embeded into $\R^3$.  But complicated manifolds could arise in condensed matter physics at least in two different situations: as subsets of $\R^3$  with complicated boundary conditions, or as configuration spaces.  Just as every orientable surface
is a polygon in the plane with glued sides, every orientable $3$-manifold can be obtained by identifying pairs of faces of a polyhedron (solid).  A familiar example is the $3$-torus as a cube with periodic boundary identifications.  Another famous example is
the Poincare homology $3$-sphere obtained by identifying faces of a dodecahedron.  Topologists use topological invariants to distinguish manifolds.   Topological invariants in physics often arise classically by integrating local geometric quantities such as curvature, or quantum mechanically by path integrals of total derivative terms, usually dropped, in the action functional.

The discovery of the fractional quantum Hall (FQH) liquids, and recently the topological insulators stirred great interests in topology in
condensed matter physics.  Witten-Chern-Simons theories as effective theories for FQH liquids and Chern numbers used in the study of topological insulators represent two kinds of topological invariants: quantum and classical.  While there is no clear-cut separation, by quantum invariants we mean  invariants of spaces obtained as path integrals of some TQFTs.  Famous examples are the
Jones polynomials at roots of unity from Witten-Chern-Simons theories.  Classical topological invariants include homotopy groups, (generalized) (co)homology groups such as $K$-theory, and characteristic classes such as Chern classes.  Using this rough division, the Jones polynomial at $4$-th root of unity appeared for the $5/2$-fractional quantum Hall effect is a quantum invariant, while the Chern number appeared in topological insulators is  a classical invariant.  But some invariants defined quantum mechanically turn out to be classically determined.

One of the frontiers in topological phases of matter is the understanding of interacting $3D$-topological insulators \cite{MQKS}\cite{SBMS}.  Interactions can be thought as
dynamical entanglement.  When entanglement becomes long-rang, topological symmetry could emerge, and an effective description by a TQFT is possible.  Therefore materialized topological symmetry can
be described by the tensor category encoding the effective TQFT.  Non-interacting topological insulators are believed to be modeled by $(3+1)$-BF TQFTs \cite{CM}, whose path integrals are determined by classical topological invariants. We ask: how many $(3+1)$-TQFTs do we know and which might be related to interacting $3D$-topological insulators?

\section{(3+1)-TQFTs}

Dimension $4$ is different.  The Euclidean space $\R^4$ is the only Euclidean space that has more than one smooth structure.  Roughly, this means that there is more than one way to do calculus on $\R^4$.  And not just one more---there are infinitely many more different ways to
do calculus on $\R^4$ \cite{Gompf}.  The classification of smooth $4$-manifolds is one of the most difficult problems in mathematics (there is another flavor of $4$-dimensional topology: the classification of topological $4$-manifolds.  Due to M.\ Freedman's work, the landscape of topological $4$-manifolds is more or less understood \cite{FreedmanQuinn}.)  Therefore, it is not surprising that there are not many smooth topological invariants, classical or quantum, in dimension $4$ that detect smooth structures.  In particular, $(3+1)$-TQFTs are very difficult to find because they could generate smooth topological invariants of $4$-manifolds.  The most powerful one for $4D$ topology is the first TQFT invented by E.\ Witten in $1988$: an $N=2$ supersymmetric Yang-Mills theory which reproduces the Donaldson invariant of $4$-manifolds \cite{Witten88}.

As a first approximation, TQFTs are quantum field theories such that the path integrals $Z(M)=Tr(e^{itH})$ for a theory with Hamiltonian $H$ on a space manifold $N$ is a topological invariant, where $M=N\times S^1$ is the space-time with periodic time.
In particular $Z(M)$ should be independent of time $t$.
$(2+1)$-Witten-Chern-Simons type TQFTs are those with $H\cong 0$, but the $(3+1)$-Witten-Donaldson TQFT is more subtle.  The trace is interpreted as a super-trace $Tr((-1)^F e^{itH})$, where $(-1)^F$ is the fermionic parity operator, so time dependence could be canceled.  Moreover, Donaldson invariant is defined only for space-time manifolds satisfying certain topological restrictions \cite{DK}.

Two interesting families $(3+1)$-TQFTs are discrete gauge theories, and $BF$ theories \cite{DW}\cite{BaezBF}.  Both families of TQFTs give rise to topological invariants that are determined by classical homotopy invariants.  The same should be true for all the $(3+1)$-TQFTs based on unitary braided fusion categories.

Witten-Donaldson theory is not determined by homotopy invariants, and detects smooth structures of $4$-manifolds.  The Witten-Donaldson TQFT is actually a partial TQFT because it is defined only for $4$-manifolds with $b_2^{+}>1$.  For example, it is not defined for the $4$-sphere $S^4$.  Very recently, Witten defined another $(3+1)$-TQFT: an $N=4$ supersymmetric Yang-Mills theory \cite{Witten11}.  It is not known if this TQFT can detect smooth structures of $4$-manifolds.  There are some other proposed $(3+1)$-TQFTs or partial TQFTs, but not much is known about them for the detection of smooth structures.

Physicists describe fermions using Grassmann numbers, which are differential $1$-forms, in the second quantized framework.
By De Rham theory, closed differential forms represent cohomology classes, thus fermions are aware of the topology of the space that they occupy.  Therefore, topological quantum field theories with fermions are potentially very different from purely bosonic ones such as $(2+1)$-Witten-Chern-Simons theories (Reshetikhin-Turaev TQFTs mathematically).  Witten-Chern-Simons theories are bosonic in the sense that the path integrals contain no Grassmann numbers.

\subsection{Discrete Gauge Theories}

Discrete gauge theories based on finite groups $G$ are the best understood examples of higher dimensional TQFTs \cite{Freed}.  For a $(3+1)$-TQFT based on $G$, the path integral of a $4$-manifold $W$ counts the number of representations from the fundamental group of $W$ to $G$.  The Hilbert space associated to a $3$-manifold $X$ is spanned by the conjugacy classes of homomorphisms from the fundamental group of $X$ to $G$.

\subsection{BF Theories}

Given a Lie group $G$, and a principle $G$-bundle $P$ over a $4$-manifold $W$, the $(3+1)$-BF theory is a TQFT based on the action
$$ S_{BF}=\int_W tr(B\wedge F+\frac{\Lambda}{12} B\wedge B),$$
where $B$ is a $adP$-valued $2$-form, $F$ the curvature of a connection $A$ on $P$, and $\Lambda$ is a coupling constant \cite{BaezBF}.  When $G=GL(4, \C)$, $\Lambda$ plays the role of the cosmological constant.  The path integral for a $4$-manifold $W$ is $e^{i \beta \sigma(W)}$, where $\beta$ is some constant independent of $W$, and $\sigma(W)$ is the signature of $W$.  The Hilbert space associated to a $3$-manifold $X$ is always $1$-dimensional.

\subsection{Cohomological Field Theories}

Witten-Donaldson type $(n+1)$-TQFT are cohomological in the sense that the vector space associated to an $n$-manifold $M$ is related to the cohomology of some moduli space.  For Donaldson theory, fix a compact Lie group $G$ and a principle $G$-bundle $P$ over a $4$-manifold $W$.  Then the moduli space is the space of anti-self dual instantons,  i.e., solution to the Yang-Mills equation $F_A^{\dagger}=0$ for connections $A$ on $P$.  Donaldson invariant roughly counts algebraically the number of solutions.  An important ingredient in the formulation of Donaldson theory is supersymmetry (topological twist, which is also very important, will not be discussed here.)  Witten formulated Donaldson invariant as correlation functions of an $N=2$ supersymmetric Yang-Mills theory \cite{Witten88}.  A beautiful illustration of the role of supersymmetry is the following formula\footnote{The second author learned the formula from J.\ Maciejko who attributed it to Witten.} to count algebraically the number $\Delta_f$ of solutions to $f(x)=0$ of a generic smooth function $f: [0,2]\rightarrow [-1,1]$ such that $f(0)=-1, f(2)=1$:
$$ \Delta_f=\int \int \int \int e^{yf(x)+\chi f'(x) \psi} dx dy d\chi d\psi,$$
where $\chi$ and $\psi$ are Grassmann variables.

An important event in $4D$ topology is the discovery of Seiberg-Witten equation around $1994$ \cite{Witten94}.  Seiberg-Witten invariant is conjectured to give the same topological information about $4$-manifolds as Donaldson invariant, and is much easier to work with mathematically.  Seiberg-Witten theory is the infra-red limit of the $N=2$ supersymmetric Yang-Mills theory.   Given a $spin^c$-structure on the $4$-manifold $W$, there associated two spinor bundles $\Lambda^{\pm}$ and a line bundle $L$ over $W$.  The Seiberg-Witten equation is for a pair $(A, \psi)$, where $A$ is connection on $L$ and $\psi$ a positive spinor:
$$ D_A \psi=0, F_A^{\dagger}=-\sigma(\psi,\psi),$$ where $D_A$ is the Dirac operator, and $\sigma$ is some paring for positive spinors.
The Seiberg-Witten invariant basically counts the number of solutions.

\section{Discrete Gauge theory coupled with anyons}

\subsection{Rigorously Solvable Hamiltoninans}

The existence of rigorously solvable Hamiltonians follows from a general mathematical theory of picture TQFTs \cite{Walker06}.  Such $(3+1)$-TQFTs based on unitary braided fusion categories will be studied mathematically in \cite{InPreparation}.  In this paper, we will write down explicit  Hamiltonians for those $(3+1)$-TQFTs and quote necessary results from \cite{Walker06}\cite{InPreparation}.

For convenience, we will describe the model on the cubic lattice in detail, and only sketch the model for general $3$-manifolds.

\subsubsection{Algebraic data}

There are several equivalent definitions of a unitary braided fusion category (UBFC).  We refer the interested readers to \cite{Wangbook}.
While every definition is complicated, there is a graphical calculus and interpretation for a UBFC as an algebraic theory of anyons
which make the axioms reasonable.  The graphs involved can be thought as anyon trajectories and interesting topological changes correspond to physical events such as creation/annihilation, fusion/spliting, braiding and twisting.   Given an anyoic system with particle types $a,b,c,...$ which form a label set $L$.  The number of particle types is finite and called the rank of the UBFC.  A particular convenient way to present a UBFC is through three sets of numbers: $\{t_a=\pm 1\}_{a\in L}$, $\{F^{abc}_{d;nm}\}, a,b,c,d,m,n\in L$, and $\{R_{c}^{ab}\}, a,b,c\in L$.  The numbers  $\{t_a=\pm 1\}_{a\in L}$ encode the creation/annihilation structures.  The so-called $6j$ symbols $\{F^{abc}_{d;nm}\}, a,b,c,d,,m,n\in L$ are solutions to the pentagon equations, while $\{R_{c}^{ab}\}, a,b,c\in L$ encode the braidings which satisfy the hexagon equations.  The three set of numbers are not independent as the creation/annihilation, fusion/spliting, and braiding have to be compatible.  (Strictly speaking the $6j$-symbols are $10j$ symbols and the braidings $R_{c}^{ab}$ are unitary matrices when some fusion multiplicity is $>1$.  For simplicity, we assume all UBFCs are multiplicity-free.)  As a consequence, any planar trivalent graph with braidings can be evaluated to a number, which is the topological amplitude of the represented physical process.

The $6j$-symbols $\{F^{abc}_{d;nm}\}$ can be organized into matrices, called $F$-matrices, by the following diagram:
\[ \xy
(0,10)*{}="A"; (4,10)*{}="B";(8,10)*{}="C"; (12,-10)*{}="D";
(4,3.3)*{}="E";(8,-3.3)*{}="F"; "A";"D" **\dir{-}; "B";"E"
**\dir{-};"C";"F" **\dir{-}; (0,12)*{a};(4,12)*{b};
(8,12)*{c};(12,-12)*{d}; (16,0)*{}="S"; (20,0)*{}="S'"; "S";"S'"
**\dir2{-};
(30,0)*{\sum_{n}F_{d;nm}^{abc}};
(48,10)*{}="A'";
(52,10)*{}="B'";(56,10)*{}="C'"; (44,-10)*{}="D'";
(48,-3.3)*{}="E'";(52,3.3)*{}="F'"; "C'";"D'" **\dir{-}; "B'";"F'"
**\dir{-};"A'";"E'" **\dir{-}; (48,12)*{a};(52,12)*{b};
(56,12)*{c};(44,-12)*{d}; (4,-3)*{m}; (52,-3)*{n};
\endxy \]

Similarly, the braiding eigenvalues are defined by the following diagram:

\[ \xy
(0,12)*{}="A"; (6,12)*{}="B"; (3,0)*{}="V"; (3,-10)*{}="C";
"V";"C" **\dir{-}; "V";"B" **\crv{(1,2) & (1,4)};
\POS?(.65)*{\hole}="2z"; "V";"2z" **\crv{(5,2) & (5,4)}; "2z";"A"
**\crv{(3,5) & (2,6)}; (0,14)*{a};(6,14)*{b};(3,-12)*{c};
(12,0)*{}="S"; (14,0)*{}="S'"; "S";"S'" **\dir2{-};
(20,0)*{R_c^{ab}}; (26,12)*{}="A'"; (32,12)*{}="B'";
(29,0)*{}="V'"; (29,-10)*{}="C'"; "V'";"C'" **\dir{-}; "V'";"A'"
**\dir{-}; "V'";"B'" **\dir{-};
(26,14)*{a};(32,14)*{b};(29,-12)*{c};
\endxy \]

Besides multiplicity-free in the fusion rules, we will also assume that all labels are self-dual, so edges in our lattices are
not oriented.  Furthermore, we assume edges around any vertex can be bent as long as we do not introduce any crossings.  Examples of such theories are premodular categories from the Temperley-Lieb algebras \cite{Wangbook}.  The Hamiltonian below generalizes to the general case with adequate notation.

\subsubsection{Cubic lattice model}

Given $\{F^{abc}_{d;nm}\}, \{R_c^{ab}\}$ of a UBFC with label set $L$ (strictly speaking, we should choose a set of representative simple objects).  Let $\C^{L}$ be the Hilbert space spanned by all labels.  Just as in the Levin-Wen model, it is convenient to work with trivalent graphs, therefore we first resolve the cubic lattice $C$ into a trivalent lattice.  There are many ways to do it, and the resulting theories are all equivalent as each one is equivalent to the continuous limit.  At each $6$-valent vertex, we resolve it as follows:

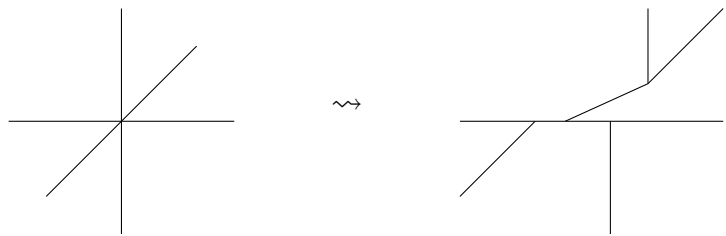
\[ \begin{tikzpicture}
\begin{scope}
\draw (-1.5,0)--(1.5,0);
\draw (0,-1.5)--(0,1.5);
\draw (-1,-1)--(1,1);
\end{scope}
\node (->) at (3,0.2) {$\rightsquigarrow$};
\begin{scope}[xshift = 6cm]
\draw (-1.5,0)--(2,0);
\draw (0.5,0)--(0.5,-1.5);
\draw (-1.5,-1)--(-0.5,0);
\draw (1,0.5)--(2,1.5);
\draw (1,0.5)--(1,1.5);
\draw (-0.1,0)--(1,0.5);
\end{scope}
\end{tikzpicture} \]

As a result, each $6$-valent vertex of the cubic lattice $C$ is split into four trivalent ones with a $\Z_3$-symmetry.  We assume periodic boundary conditions, so our lattice is in the $3$-torus.

The Hilbert space $V$ of the model is spanned by all labelings of the edges in the resolved cubic lattice, denoted by $\Gamma_C$.   Equivalently we assign a qudit $\C^L$ to each edge of $\Gamma_C$ including the new ones.  There are two kinds of terms in the Hamiltonian: the vertex type and the plaquette type.  In order to write down the Hamiltonian explicitly, we need to project our lattice onto a plane.  We look at the lattice along the $(1,1,1)$-direction, and therefore the resolved cubic lattice $\Gamma_C$ is projected to a plane perpendicular to the $(1,1,1)$-direction.  The microscopic Hilbert space is $V=\otimes_{e\in \Gamma_C} \C^L$.  The Hamiltonian $H$ is of the form $-J_1\sum_{v\in \Gamma_C}A_v -J_2\sum_{\p\in C_{sq}} B_\p$, where $v$ ranges over all vertices of $\Gamma_C$ including the new ones, $\p$ ranges over $C_{sq}$---all plaquettes which correspond to the original squares of the cubic lattice $C$, and $J_1, J_2$ are coupling constants.

The microscopic Hilbert space $V=\otimes_{e\in \Gamma} \C^L$ has a natural basis: the $1$-skeleton of $\Gamma_C$ with a label on each edge.  Then for a vertex $v$ of $\Gamma_C$, the term $A_v$ acts a basis $|\Psi\rangle$ as follows: $A_v|\Psi\rangle=|\Psi\rangle$ if the three labels around $v$ obey the fusion rules; otherwise $A_v|\Psi\rangle=0$.  A plaquette term is much more complicated, we explain below how it is derived and give its formula in terms of $6j$-symbols and braiding eigenvalues.

Our resolution of the cubic lattice to a trivalent one is very symmetric.  There are three kinds of plaquettes in $C_{sq}$: those corresponding to squares in the planes parallel to $x$-$y$, $y$-$z$ and $x$-$z$ plane.  Below are the pictures of three representatives of such plaquettes.

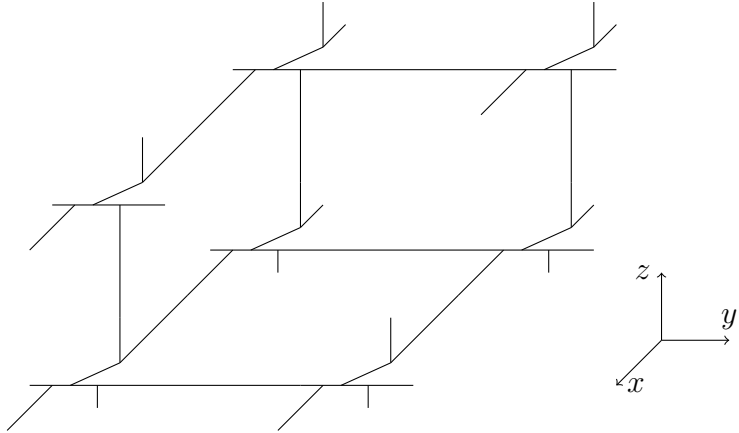
\[ \begin{tikzpicture}[scale=0.6]
\begin{scope}
\draw (-1,0)--(5,0);
\draw (0.5,0)--(0.5,-0.5);
\draw (-1.5,-1)--(-0.5,0);
\draw (1,0.5)--(2.5,2);
\draw (1,0.5)--(1,1.5);
\draw (-0.1,0)--(1,0.5);
\end{scope}
\begin{scope}[xshift=6cm]
\draw (-1,0)--(1.5,0);
\draw (0.5,0)--(0.5,-0.5);
\draw (-1.5,-1)--(-0.5,0);
\draw (1,0.5)--(2.5,2);
\draw (1,0.5)--(1,1.5);
\draw (-0.1,0)--(1,0.5);
\end{scope}
\begin{scope}[xshift=4cm,yshift=3cm]
\draw (-1,0)--(5,0);
\draw (0.5,0)--(0.5,-0.5);
\draw (-1.5,-1)--(-0.5,0);
\draw (1,0.5)--(1.5,1);
\draw (1,0.5)--(1,1.5);
\draw (-0.1,0)--(1,0.5);
\end{scope}
\begin{scope}[xshift=10cm,yshift=3cm]
\draw (-1,0)--(1.5,0);
\draw (0.5,0)--(0.5,-0.5);
\draw (-1.5,-1)--(-0.5,0);
\draw (1,0.5)--(1.5,1);
\draw (1,0.5)--(1,1.5);
\draw (-0.1,0)--(1,0.5);
\end{scope}
\begin{scope}[xshift=0.5cm,yshift=4cm]
\draw (-1,0)--(1.5,0);
\draw (0.5,0)--(0.5,-2.5);
\draw (-1.5,-1)--(-0.5,0);
\draw (1,0.5)--(2,1.5);
\draw (1,0.5)--(1,1.5);
\draw (-0.1,0)--(1,0.5);
\end{scope}
\begin{scope}[xshift=4.5cm,yshift=7cm]
\draw (-1,0)--(4.5,0);
\draw (0.5,0)--(0.5,-2.5);
\draw (-2,-1.5)--(-0.5,0);
\draw (1,0.5)--(1.5,1);
\draw (1,0.5)--(1,1.5);
\draw (-0.1,0)--(1,0.5);
\end{scope}
\begin{scope}[xshift=10.5cm,yshift=7cm]
\draw (-1.5,0)--(1.5,0);
\draw (0.5,0)--(0.5,-2.5);
\draw (-1.5,-1)--(-0.5,0);
\draw (1,0.5)--(1.5,1);
\draw (1,0.5)--(1,1.5);
\draw (-0.1,0)--(1,0.5);
\end{scope}
\begin{scope}[xshift=13cm,yshift=1cm]
\draw [->](0,0)--(1.5,0);
\draw [->](0,0)--(0,1.5);
\draw [->](0,0)--(-1,-1);
\draw (-1,-1) node[anchor=west] {$x$};
\draw (1.5,0) node[anchor=south] {$y$};
\draw (0,1.5) node[anchor=east] {$z$};
\end{scope}
\end{tikzpicture} \]

Note that each plaquette $\p$ is a decagon (although we will call such a plaquette a decagon, it is not always planar.  It is a surface bounded by a polygonal path with $10$ edges: $4$ of them correspond to the $4$ sides of the original square of the cubic lattice and $6$ new ones from the resolution.)  There are also $10$ adjacent edges.  The plaquette term indexed by $\p$ is obtained by placing a simple loop labeled by a  projector $\omega_0$ onto the plaquette along the edges as below.

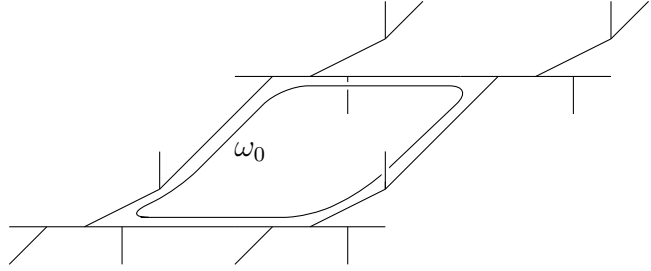
\[ \begin{tikzpicture}[scale=0.5]
\begin{scope}
\draw (-2,0)--(4,0);
\draw (1,0)--(1,-1);
\draw (-2,-1)--(-1,0);
\draw (2,1)--(4,3);
\draw (2,1)--(2,2);
\draw (0,0)--(2,1);
\end{scope}
\begin{scope}[xshift=6cm]
\draw (-2,0)--(2,0);
\draw (1,0)--(1,-1);
\draw (-2,-1)--(-1,0);
\draw (2,1)--(4,3);
\draw (2,1)--(2,2);
\draw (0,0)--(2,1);
\end{scope}
\begin{scope}[xshift=6cm,yshift=4cm]
\draw (-2,0)--(4,0);
\draw (1,0)--(1,-1);
\draw (-2,-1)--(-1,0);
\draw (2,1)--(3,2);
\draw (2,1)--(2,2);
\draw (0,0)--(2,1);
\end{scope}
\begin{scope}[xshift=12cm,yshift=4cm]
\draw (-2,0)--(2,0);
\draw (1,0)--(1,-1);
\draw (-2,-1)--(-1,0);
\draw (2,1)--(3,2);
\draw (2,1)--(2,2);
\draw (0,0)--(2,1);
\end{scope}
\draw [color=white,line width=1mm] (10.4,3.75)--(5.25,3.75);
\draw [rounded corners=10pt](1.5,0.25)--(6,0.25)--(7.5,1)--(10.4,3.75)--(5.25,3.75)--(2.5,1)--(1,0.25)--(1.5,0.25);
\draw [color=white,line width=1mm] (8,1.2)--(8,2);
\draw (8,1)--(8,2);
\draw (3.7,2) node[anchor=west] {$\omega_0$};
\end{tikzpicture} \]

Physically the projector $\omega_0$ enforces the total flux through $\p$ to be a transparent label (see Section \ref{mathpinning} below for the definition).  Such a projector is formally written as $\omega_0=\sum_{s\in L}\frac{d_s}{D^2} s$, where $d_s$ is the quantum dimension of the label $s$ and $D^2=\sum_{s\in L}d_s^2$.  Adding such a loop with the projector $\omega_0$ will not change the topological amplitude of a basis $|\Psi\rangle$.  This can be seen by expanding $\omega_0$ into $\sum_{s\in L}\frac{d_s}{D^2} s$ and noticing that a contractible loop labeled by $s$ is evaluated to $d_s$.  A formula for a plaquette term is then obtained by evaluating the same projector in a different way using $6j$ symbols and braiding eigenvalues.  For analogous derivations of similar terms in Levin-Wen model, see page $100$ of \cite{Wangbook}.

Due to the regularity of the cubic lattice and the symmetry of our resolution, we need only to write down one plaquette term.  We choose to write the formula for the plaquette in the $x$-$y$-plane, denoted as $\p_{xy}$.  We could equally work with the one in the $x$-$z$-plane $\p_{xz}$ or the one in the $y$-$z$-plane $\p_{yz}$.

To write down such a formula, we denote the basis element that labels the $10$ edges of $\p_{xy}$ by $abcdpqruvw$ and their $10$ adjacent edges by $a'b'c'd'p'q'r'u'v'w'$ as in the following picture by $|\Psi_{\p_{xy},abcdpqruvw}\rangle$.  Labels of edges that are not named remain the same in all computations.  Our convention is that the edge not on the decagon $\p_{xy}$, but next to the edge of the decagon labeled by $l$, is labeled by $l'$.

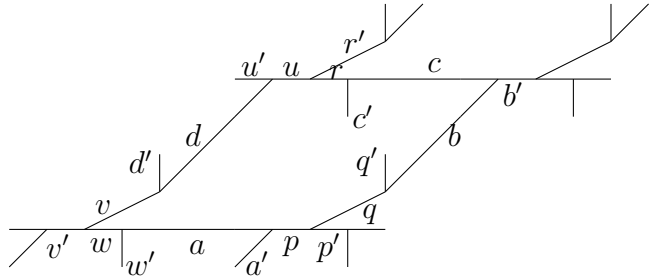
\[ \begin{tikzpicture}[scale=.5]
\begin{scope}
\draw (-2,0)--(4,0);
\draw (1,0)--(1,-1);
\draw (-2,-1)--(-1,0);
\draw (2,1)--(4,3);
\draw (2,1)--(2,2);
\draw (0,0)--(2,1);
\end{scope}
\begin{scope}[xshift=6cm]
\draw (-2,0)--(2,0);
\draw (1,0)--(1,-1);
\draw (-2,-1)--(-1,0);
\draw (2,1)--(4,3);
\draw (2,1)--(2,2);
\draw (0,0)--(2,1);
\end{scope}
\begin{scope}[xshift=6cm,yshift=4cm]
\draw (-2,0)--(4,0);
\draw (1,0)--(1,-1);
\draw (-2,-1)--(-1,0);
\draw (2,1)--(3,2);
\draw (2,1)--(2,2);
\draw (0,0)--(2,1);
\end{scope}
\begin{scope}[xshift=12cm,yshift=4cm]
\draw (-2,0)--(2,0);
\draw (1,0)--(1,-1);
\draw (-2,-1)--(-1,0);
\draw (2,1)--(3,2);
\draw (2,1)--(2,2);
\draw (0,0)--(2,1);
\end{scope}
\draw (3,0) node[anchor=north] {$a$};
\draw (5.2,-0.9) node[anchor=east] {$a'$};
\draw (5.5,0.1) node[anchor=north] {$p$};
\draw (6.5,0.2) node[anchor=north] {$p'$};
\draw (7.1,0.4) node[anchor=west] {$q$};
\draw (8.1,1.8) node[anchor=east] {$q'$};
\draw (10.3,2.5) node[anchor=east] {$b$};
\draw (11.4,4.2) node[anchor=north] {$b'$};
\draw (9.3,3.9) node[anchor=south] {$c$};
\draw (7.4,2.4) node[anchor=south] {$c'$};
\draw (6.7,3.7) node[anchor=south] {$r$};
\draw (6.6,5) node[anchor=west] {$r'$};
\draw (5.5,3.8) node[anchor=south] {$u$};
\draw (4.5,3.8) node[anchor=south] {$u'$};
\draw (3.4,2.5) node[anchor=east] {$d$};
\draw (2.1,1.8) node[anchor=east] {$d'$};
\draw (1,0.6) node[anchor=east] {$v$};
\draw (-0.7,0.15) node[anchor=north] {$v'$};
\draw (0.45,0.05) node[anchor=north] {$w$};
\draw (0.8,-0.9) node[anchor=west] {$w'$};
\end{tikzpicture} \]

The plaquette term $B_{\p_{xy}}$ will map the basis vector $|\Psi_{{\p_{xy}},abcdpqruvw}\rangle$ into a big linear combination of basis elements, where the labels $a,b,c,d,p,q,r,u,v,w$ are replaced by new labels $a'',b'',c'',d'',p'',q'',r'',u'',v'',w''$.  In the following, $$a'',b'',c'',d'',p'',q'',r'',u'',v'',w''$$ will be abbreviated as $a'',...,w''$.  Therefore, all we need are the coefficients $B^s_{\p_{xy},a'',...,w''}$ in $B_{\p_{xy}} |\Psi_{{\p_{xy}},abcdpqruvw}\rangle=$
$$\sum_{s\in L} \frac{d_s}{D^2} \sum_{a'',...,w''\in L} B^s_{\p_{xy},a'',...,w''} |\Psi^s_{\p_{xy},a'',...,w''}\rangle.$$

 Recall that $\omega_0$ is the formal sum $\sum_{s\in L}\frac{d_s}{D^2} s$.  The operator $B_{\p_{xy}}=\sum_{s\in L}\frac{d_s}{D^2} B^s_{\p_{xy}}$ is a sum of operators $B^s_{\p_{xy}}$, where $B^s_{\p_{xy}}$ is the operator that corresponds to the simple loop labeled by $s$.  Hence it suffices to know the coefficients $B^s_{{\p_{xy}},a'',...,w''}$ in $B^s_{\p_{xy}}|\Psi_{{\p_{xy}},abcdpqruvw}\rangle=\sum_{a'',...,w''}B^s_{{\p_{xy}},a'',...,w''} |\Psi^s_{{\p_{xy}},a'',...,w''}\rangle.$  In terms of $6j$ symbols and braiding eigenvalues, we claim

$B^s_{{\p_{xy}},a'',...,w''}=$
$$R_{q}^{q'b}\overline{R_{c}^{c'r}}\overline{R_{q''}^{q'b''}}R_{c''}^{c'r''}F_{a';ap''}^{a''sp} F_{p';pq''}^{p''sq} F_{q';qb''}^{q''sb} F_{b';bc''}^{b''sc} F_{c';cr''}^{c''sr} F_{r';ru''}^{r''su} F_{u';ud''}^{u''sd} F_{d';dv''}^{d''sv} F_{v';vw''}^{v''sw} F_{w';wa''}^{w''sa}.$$

To derive this formula, we first twist the labeled graph representing the basis $|\Psi_{{\p_{xy}},abcdxyzuvw}\rangle$ around the two vertical edges as below.

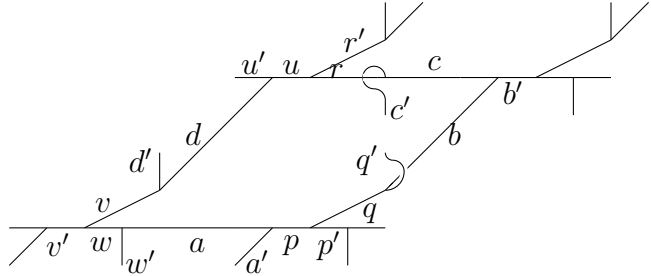
\[ \begin{tikzpicture}[scale=.5]
\begin{scope}
\draw (-2,0)--(4,0);
\draw (1,0)--(1,-1);
\draw (-2,-1)--(-1,0);
\draw (2,1)--(4,3);
\draw (2,1)--(2,2);
\draw (0,0)--(2,1);
\end{scope}
\begin{scope}[xshift=6cm]
\draw (-2,0)--(2,0);
\draw (1,0)--(1,-1);
\draw (-2,-1)--(-1,0);
\draw (2,1)--(4,3);
\draw (0,0)--(2,1);
\end{scope}
\begin{scope}[xshift=6cm,yshift=4cm]
\draw (-2,0)--(4,0);
\draw (-2,-1)--(-1,0);
\draw (2,1)--(3,2);
\draw (2,1)--(2,2);
\draw (0,0)--(2,1);
\end{scope}
\begin{scope}[xshift=12cm,yshift=4cm]
\draw (-2,0)--(2,0);
\draw (1,0)--(1,-1);
\draw (-2,-1)--(-1,0);
\draw (2,1)--(3,2);
\draw (2,1)--(2,2);
\draw (0,0)--(2,1);
\end{scope}
\draw (3,0) node[anchor=north] {$a$};
\draw (5.2,-0.9) node[anchor=east] {$a'$};
\draw (5.5,0.1) node[anchor=north] {$p$};
\draw (6.5,0.2) node[anchor=north] {$p'$};
\draw (7.1,0.4) node[anchor=west] {$q$};
\draw (8.1,1.8) node[anchor=east] {$q'$};
\draw (10.3,2.5) node[anchor=east] {$b$};
\draw (11.4,4.2) node[anchor=north] {$b'$};
\draw (9.3,3.9) node[anchor=south] {$c$};
\draw (8.4,2.6) node[anchor=south] {$c'$};
\draw (6.7,3.7) node[anchor=south] {$r$};
\draw (6.6,5) node[anchor=west] {$r'$};
\draw (5.5,3.8) node[anchor=south] {$u$};
\draw (4.5,3.8) node[anchor=south] {$u'$};
\draw (3.4,2.5) node[anchor=east] {$d$};
\draw (2.1,1.8) node[anchor=east] {$d'$};
\draw (1,0.6) node[anchor=east] {$v$};
\draw (-0.7,0.15) node[anchor=north] {$v'$};
\draw (0.45,0.05) node[anchor=north] {$w$};
\draw (0.8,-0.9) node[anchor=west] {$w'$};
\draw [color=white,line width=1mm](8,1) arc(270:360:0.5);
\draw [color=white,line width=1mm](8.5,1.5) arc(0:90:0.3);
\draw [color=white,line width=1mm](8.2,1.8) arc(270:180:0.2);
\draw (8,1)--(8.2,1.2);
\draw (8,1) arc(270:360:0.5);
\draw (8.5,1.5) arc(0:90:0.3);
\draw (8.2,1.8) arc(270:180:0.2);
\draw (8,4) arc(0:270:0.3);
\draw (7.7,3.7) arc(90:0:0.3);
\draw [color=white,line width=1mm](7.4,4)--(7.8,4);
\draw (7.4,4)--(7.8,4);
\draw (8,3.4)--(8,3);
\end{tikzpicture} \]

This multiplies $|\Psi_{p,abcdxyzuvw}\rangle$ by $\overline{R_{q}^{q'b}} R_{u}^{u'c}$.  Then we fuse the simple loop labeled by $s$ with the edge labeled by $a$ as shown below.

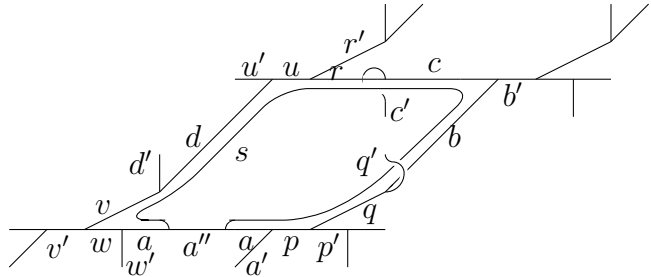
\[ \begin{tikzpicture}[scale=.5]
\begin{scope}
\draw (-2,0)--(4,0);
\draw (1,0)--(1,-1);
\draw (-2,-1)--(-1,0);
\draw (2,1)--(4,3);
\draw (2,1)--(2,2);
\draw (0,0)--(2,1);
\end{scope}
\begin{scope}[xshift=6cm]
\draw (-2,0)--(2,0);
\draw (1,0)--(1,-1);
\draw (-2,-1)--(-1,0);
\draw (2,1)--(4,3);
\draw (0,0)--(2,1);
\end{scope}
\begin{scope}[xshift=6cm,yshift=4cm]
\draw (-2,0)--(4,0);
\draw (-2,-1)--(-1,0);
\draw (2,1)--(3,2);
\draw (2,1)--(2,2);
\draw (0,0)--(2,1);
\end{scope}
\begin{scope}[xshift=12cm,yshift=4cm]
\draw (-2,0)--(2,0);
\draw (1,0)--(1,-1);
\draw (-2,-1)--(-1,0);
\draw (2,1)--(3,2);
\draw (2,1)--(2,2);
\draw (0,0)--(2,1);
\end{scope}
\draw (3,0.2) node[anchor=north] {$a''$};
\draw (5.2,-0.9) node[anchor=east] {$a'$};
\draw (5.5,0.1) node[anchor=north] {$p$};
\draw (6.5,0.2) node[anchor=north] {$p'$};
\draw (7.1,0.4) node[anchor=west] {$q$};
\draw (8.1,1.8) node[anchor=east] {$q'$};
\draw (10.3,2.5) node[anchor=east] {$b$};
\draw (11.4,4.2) node[anchor=north] {$b'$};
\draw (9.3,3.9) node[anchor=south] {$c$};
\draw (8.4,2.6) node[anchor=south] {$c'$};
\draw (6.7,3.7) node[anchor=south] {$r$};
\draw (6.6,5) node[anchor=west] {$r'$};
\draw (5.5,3.8) node[anchor=south] {$u$};
\draw (4.5,3.8) node[anchor=south] {$u'$};
\draw (3.4,2.5) node[anchor=east] {$d$};
\draw (2.1,1.8) node[anchor=east] {$d'$};
\draw (1,0.6) node[anchor=east] {$v$};
\draw (-0.7,0.15) node[anchor=north] {$v'$};
\draw (0.45,0.05) node[anchor=north] {$w$};
\draw (0.8,-0.9) node[anchor=west] {$w'$};
\draw (8,4) arc(0:270:0.3);
\draw (7.7,3.7) arc(90:0:0.3);
\draw [color=white,line width=1mm](7.4,4)--(7.8,4);
\draw (7.4,4)--(7.8,4);
\draw (8,3.4)--(8,3);
\draw [color=white,line width=1.5mm] (10.4,3.75)--(5.25,3.75);
\draw [rounded corners=10pt](1.5,0.25)--(6,0.25)--(7.5,1)--(10.4,3.75)--(5.25,3.75)--(2.5,1)--(1,0.25)--(1.5,0.25);
\draw (3.7,2) node[anchor=west] {$s$};
\draw (2,0.25) arc(90:0:0.25);
\draw (4,0.25) arc(90:180:0.25);
\draw [color=white,line width=0.5mm](2.15,0.25)--(3.85,0.25);
\draw (1.6,0.1) node[anchor=north] {$a$};
\draw (4.3,0.1) node[anchor=north] {$a$};
\draw [color=white,line width=1mm](8,1) arc(270:360:0.5);
\draw [color=white,line width=1mm](8.5,1.5) arc(0:90:0.3);
\draw [color=white,line width=1mm](8.2,1.8) arc(270:180:0.2);
\draw (8,1)--(8.2,1.2);
\draw (8,1) arc(270:360:0.5);
\draw (8.5,1.5) arc(0:90:0.3);
\draw (8.2,1.8) arc(270:180:0.2);
\end{tikzpicture} \]

Next a sequence of $F$-moves brings the $s$-labeled strand counter-clock-wise along the boundary of the decagon $p$ through all the trivalent vertices one by one.  Each time when the $s$-labeled strand passes a trivalent vertex on $p$, an $F$-move is used.  Due to the two introduced twists, we do not need to use braidings when we perform all the $F$-moves.

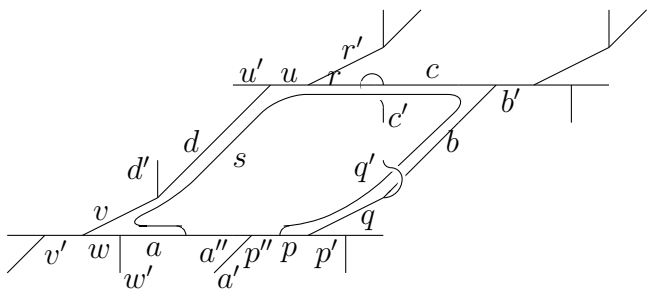
\[ \begin{tikzpicture}[scale=.5]
\begin{scope}
\draw (-2,0)--(4,0);
\draw (1,0)--(1,-1);
\draw (-2,-1)--(-1,0);
\draw (2,1)--(4,3);
\draw (2,1)--(2,2);
\draw (0,0)--(2,1);
\end{scope}
\begin{scope}[xshift=6cm]
\draw (-2,0)--(2,0);
\draw (1,0)--(1,-1);
\draw (-2.5,-1)--(-1.5,0);
\draw (2,1)--(4,3);
\draw (0,0)--(2,1);
\end{scope}
\begin{scope}[xshift=6cm,yshift=4cm]
\draw (-2,0)--(4,0);
\draw (-2,-1)--(-1,0);
\draw (2,1)--(3,2);
\draw (2,1)--(2,2);
\draw (0,0)--(2,1);
\end{scope}
\begin{scope}[xshift=12cm,yshift=4cm]
\draw (-2,0)--(2,0);
\draw (1,0)--(1,-1);
\draw (-2,-1)--(-1,0);
\draw (2,1)--(3,2);
\draw (2,1)--(2,2);
\draw (0,0)--(2,1);
\end{scope}
\draw (3.5,0.2) node[anchor=north] {$a''$};
\draw (4.5,-1.1) node[anchor=east] {$a'$};
\draw (5.5,0.1) node[anchor=north] {$p$};
\draw (6.5,0.2) node[anchor=north] {$p'$};
\draw (7.1,0.4) node[anchor=west] {$q$};
\draw (8.1,1.8) node[anchor=east] {$q'$};
\draw (10.3,2.5) node[anchor=east] {$b$};
\draw (11.4,4.2) node[anchor=north] {$b'$};
\draw (9.3,3.9) node[anchor=south] {$c$};
\draw (8.4,2.6) node[anchor=south] {$c'$};
\draw (6.7,3.7) node[anchor=south] {$r$};
\draw (6.6,5) node[anchor=west] {$r'$};
\draw (5.5,3.8) node[anchor=south] {$u$};
\draw (4.5,3.8) node[anchor=south] {$u'$};
\draw (3.4,2.5) node[anchor=east] {$d$};
\draw (2.1,1.8) node[anchor=east] {$d'$};
\draw (1,0.6) node[anchor=east] {$v$};
\draw (-0.7,0.15) node[anchor=north] {$v'$};
\draw (0.45,0.05) node[anchor=north] {$w$};
\draw (0.8,-1.1) node[anchor=west] {$w'$};
\draw (8,4) arc(0:270:0.3);
\draw (7.7,3.7) arc(90:0:0.3);
\draw [color=white,line width=1mm](7.4,4)--(7.8,4);
\draw (7.4,4)--(7.8,4);
\draw (8,3.4)--(8,3);
\draw [color=white,line width=1.5mm] (10.4,3.75)--(5.25,3.75);
\draw [rounded corners=10pt](1.5,0.25)--(6,0.25)--(7.5,1)--(10.4,3.75)--(5.25,3.75)--(2.5,1)--(1,0.25)--(1.5,0.25);
\draw (3.7,2) node[anchor=west] {$s$};
\draw (2.5,0.25) arc(90:0:0.25);
\draw (5.5,0.25) arc(90:180:0.25);
\draw [color=white,line width=0.5mm](2.65,0.25)--(5.35,0.25);
\draw (4.7,0.2) node[anchor=north] {$p''$};
\draw (1.9,0.1) node[anchor=north] {$a$};
\draw [color=white,line width=1mm](8,1) arc(270:360:0.5);
\draw [color=white,line width=1mm](8.5,1.5) arc(0:90:0.3);
\draw [color=white,line width=1mm](8.2,1.8) arc(270:180:0.2);
\draw (8,1)--(8.2,1.2);
\draw (8,1) arc(270:360:0.5);
\draw (8.5,1.5) arc(0:90:0.3);
\draw (8.2,1.8) arc(270:180:0.2);
\end{tikzpicture} \]

Finally, the $s$-labeled strand returns to the edge labeled by $a$ with a bubble.

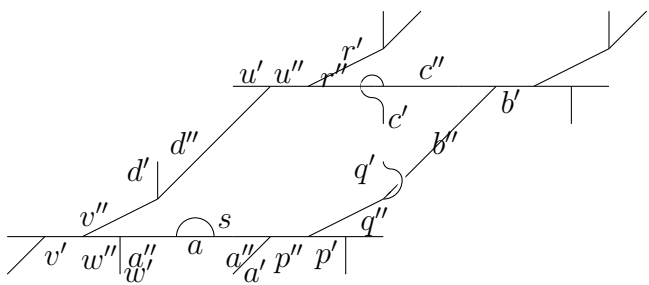
\[ \begin{tikzpicture}[scale=.5]
\begin{scope}
\draw (-2,0)--(4,0);
\draw (1,0)--(1,-1);
\draw (-2,-1)--(-1,0);
\draw (2,1)--(4,3);
\draw (2,1)--(2,2);
\draw (0,0)--(2,1);
\end{scope}
\begin{scope}[xshift=6cm]
\draw (-2,0)--(2,0);
\draw (1,0)--(1,-1);
\draw (-2,-1)--(-1,0);
\draw (2,1)--(4,3);
\draw (0,0)--(2,1);
\end{scope}
\begin{scope}[xshift=6cm,yshift=4cm]
\draw (-2,0)--(4,0);
\draw (-2,-1)--(-1,0);
\draw (2,1)--(3,2);
\draw (2,1)--(2,2);
\draw (0,0)--(2,1);
\end{scope}
\begin{scope}[xshift=12cm,yshift=4cm]
\draw (-2,0)--(2,0);
\draw (1,0)--(1,-1);
\draw (-2,-1)--(-1,0);
\draw (2,1)--(3,2);
\draw (2,1)--(2,2);
\draw (0,0)--(2,1);
\end{scope}
\draw (3,0.2) node[anchor=north] {$a$};
\draw (5.2,-0.9) node[anchor=east] {$a'$};
\draw (5.5,0.1) node[anchor=north] {$p''$};
\draw (6.5,0.2) node[anchor=north] {$p'$};
\draw (7.1,0.4) node[anchor=west] {$q''$};
\draw (8.1,1.8) node[anchor=east] {$q'$};
\draw (10.3,2.5) node[anchor=east] {$b''$};
\draw (11.4,4.2) node[anchor=north] {$b'$};
\draw (9.3,3.9) node[anchor=south] {$c''$};
\draw (8.4,2.6) node[anchor=south] {$c'$};
\draw (6.7,3.7) node[anchor=south] {$r''$};
\draw (6.6,5) node[anchor=west] {$r'$};
\draw (5.5,3.8) node[anchor=south] {$u''$};
\draw (4.5,3.8) node[anchor=south] {$u'$};
\draw (3.4,2.5) node[anchor=east] {$d''$};
\draw (2.1,1.8) node[anchor=east] {$d'$};
\draw (1,0.6) node[anchor=east] {$v''$};
\draw (-0.7,0.15) node[anchor=north] {$v'$};
\draw (0.45,0.05) node[anchor=north] {$w''$};
\draw (0.8,-0.9) node[anchor=west] {$w'$};
\draw [color=white,line width=1mm](8,1) arc(270:360:0.5);
\draw [color=white,line width=1mm](8.5,1.5) arc(0:90:0.3);
\draw [color=white,line width=1mm](8.2,1.8) arc(270:180:0.2);
\draw (8,1)--(8.2,1.2);
\draw (8,1) arc(270:360:0.5);
\draw (8.5,1.5) arc(0:90:0.3);
\draw (8.2,1.8) arc(270:180:0.2);
\draw (8,4) arc(0:270:0.3);
\draw (7.7,3.7) arc(90:0:0.3);
\draw [color=white,line width=1mm](7.4,4)--(7.8,4);
\draw (7.4,4)--(7.8,4);
\draw (8,3.4)--(8,3);
\draw (3.5,0) arc(0:180:0.5);
\draw (3.3 ,0.4) node[anchor=west] {$s$};
\draw (1.6,0.1) node[anchor=north] {$a''$};
\draw (4.2,0.1) node[anchor=north] {$a''$};
\end{tikzpicture} \]

Removing the bubble and twisting back the two vertical edges, we arrive at the formula above.

\subsubsection{Dual cellulations}

The cubic lattice is so regular that a lot of technical difficulties disappeared.  Since there is substantial topology involved for the general case, we are content with a sketch of the procedures.  For the basic topology involved, see the book \cite{Spine}.

Trivalent graphs are generic in surfaces under perturbations.  Genetic 2-dimensional polyhedra are simple polyhedra which are generalizations of trivalent graphs.  For a general 3-manifold $X$, our Hamiltonian is given on a branched standard spine of $X$.
An elementary way to present a $3$-manifold $X$ is by a triangulation of $X$.  A triangulation $\Delta$ of a $3$-manifold $X$ is a collection of tetrahedra $\{\Delta_i\}$ such that tetrahedra are glued together along their faces.  Our model is conveniently defined using the dual cellulation $\Gamma_{\Delta}$ of the triangulation $\Delta$.  In two dimension, recall that in the dual cellulation of a triangulation of a surface, each triangle becomes a vertex, an edge still an edge, while a vertex becomes a cell (polygon).  Notice that the $1$-skeleton of the dual cellulation of a surface is always a trivalent graph.  In $3D$, a vertex of the dual cellulation $\Gamma_{\Delta}$ is the center of a tetrahedron.  Two vertices of $\Gamma_{\Delta}$ are connected by an edge if their corresponding tetrahedra share a face.  A face of $\Gamma_{\Delta}$ is dual to an edge of $\Delta$, and a $3$-cell (solid) dual to a vertex of $\Delta$.  The dual cellutation of a 3-manifold is an example of a simple polyhedron.  Our model can be easily generalized to any branched standard spine of a $3$-manifold $X$, which is much more convenient to work with in practice.

In order to define our Hamiltonian, we need an extra structure, called an oriented branching,  on the triangulation $\Delta$, or equivalently on $\Gamma_{\Delta}$.  We will define it for the triangulation $\Delta$, but the translation to $\Gamma_{\Delta}$ is straightforward.
An oriented branching on a triangulation $\Delta$ of $X$ is an assignment of an arrow to each edge of $\Delta$ so that in the three edges of each triangle of $\Delta$ there are exactly two consistent arrows, i.e., the three arrows of any triangle never form a cycle.  Every oriented $3$-manifold $X$ has a triangulation with an oriented branching \cite{Spine}.  An oriented branching uniquely determines an ordering of the $4$ vertices of each tetrahedron if the arrows go from lower numbered vertices to higher numbered ones.

Using an oriented branching, we can identify each tetrahedron in a triangulation $\Delta$ of $X$ with the standard tetrahedron in $\R^3$.  The $1$-skeleton of the dual cellulation is a $4$-valent graph.  We can resolve this $4$-valent graph into a trivalent graph in a standard way.
The Hilbert space of our model is again the tensor product of all qudits $\C^L$ over all edges of the resolved $1$-skeleton of the dual cellulation $\Gamma_{\Delta}$ including the new ones.  There are again two kinds of terms: vertex and plaquette types.  The Hamiltonian is $H=-\sum_{v\in \Gamma}A_v -\sum_{p\in \Gamma} B_p$, where $v$ ranges over all vertices including the new ones, and $p$ ranges over all $2$-cells of the dual cellulation (plaquettes).  The oriented branching allows us to have standard local models of the standard spine in $\R^3$, therefore we can use a similar procedure as for the cubic lattice to write down the plaquette terms.

\subsection{Ground State Manifold}

Given an oriented $3$-manifold $X$, consider the infinite dimensional vector space $\tilde{A}(X)$ generated by all colored string-nets in $X$.  Let $A(X)$ be the quotient space of $\tilde{A}(X)$ by all local relations derived from the UBFC.  (Strictly speaking, the string-net strands are ribbons and the vertices are rigid.  We refer the interested readers to \cite{InPreparation} for a mathematical discussion.)  Then $A(X)$, called the skein spaces, are isomorphic to the ground state manifolds of the spin Hamiltonians, which are examples of error-correction codes.

\subsection{Statistics of Excitations}

Elementary excitations in the spin models include pointed particles, loop-defects, $\theta$-defects, and more general defects.  Their types and statistics can be described using representations of certain cylindrical categories related to boundary conditions for $3$-manifolds.

 We consider two kinds of boundary conditions for a $3$-manifold $X$ with a boundary surface $Y$: crude and topological.  A crude boundary condition $c$ on $Y$ is a finite collection of points labeled by objects of $\mC$.  Let $\tilde{A}(X,Y;c)$ be the vector space generated by all colored string-nets in $X$ which terminate at $c$.  Then the relative skein space $A(X,Y;c)$ is the quotient of $\tilde{A}(X,Y;c)$ by local relations.  We  define a category $A(Y)$ for each surface $Y$ (including the empty one) as follows.  An object of $A(Y)$ is a crude boundary condition $c$.  The morphism space from a boundary condition $c_1$ to $c_2$ is the relative skein space $A(Y\times [0,1];c_1,c_2)$.  Therefore, a morphism from an object $c_1$ to another object $c_2$ is represented by a linear combination of string-nets in $Y\times [0,1]$ that terminate at $c_1$ and $c_2$ in $Y\times 0$ and $Y\times 1$, respectively.

 Just as algebras, the linear categories $A(Y)$ have representations.  Any representation of $A(Y)$ is called a topological boundary condition.  An  irreducible representation corresponds to an elementary excitation whose boundary is $Y$.  The statistics of elementary excitations can be computed by using the relative skein space with topological boundary conditions.  $(2+1)$-dimensional analogues are discussed in \cite{FNWW}.

\subsection{Modular Category}

The $SU(2)$-Witten-Chern-Simons theories can be promoted to $(3+1)$-TQFTs, called Crane-Yetter TQFTs \cite{CraneYetter}.  The Crane-Yetter  $(3+1)$-TQFTs and their generalizations based on premodular categories induce representations of the motion groups---generalizations of the braid group to more general mapping class groups.

The generalization of the braid group to extended objects in $3D$ is conceptually straightforward.  The generalization to small loops is called the loop braid group \cite{Lin}.  Consider finitely many small $3$-balls $\{B^3_i\}$ inside the standard $3$-ball $B^3$.  Fix the equator $S_i$ of each small $3$-ball $B^3_i$.  Then the mapping class group of $B^3\backslash \{S_i\}$---the self-diffeomorphisms of $B^3\backslash \{S_i\}$ fixing the outside boundary of $B^3$ modulo deformations, is the loop braid group.  Given a group, it is not always easy to find a presentation, i.e., a set of generators and relations.  For the loop braid group, this is done in \cite{Lin}.

 Suppose $(V,Z)$ is a $(3+1)$-TQFT based on a modular category $\mC$.  For an oriented closed $4$-manifold $W$, the topological invariant $Z(W)$ is $e^{i \frac{\pi }{4} c \sigma(W)}$, where $c$ is the topological central charge of $\mC$, and $\sigma(W)$ is the signature of $W$.  For an oriented closed $3$-manifold, the vector space is always $1$-dimensional.  When an oriented $3$-manifold $X$ has a boundary $Y$, then the topological boudnary condition for $Y$ is always trivial.  The vector space associated to $X$ with this trivial boundary condition on $Y$ is isomorphic to $V_{\textrm{RT}}(Y)$---the vector space associated to $Y$ in the Reshetikhin-Turaev $(2+1)$-TQFT based on $\mC$.

 Mathematically, the category $A(Y)$ is trivial up to Morita equivalence \cite{InPreparation}.  Consequently for any oriented closed surface $Y$, there are neither non-trivial particle excitations nor non-trivial excitations of extended objects.  This is analogous to the situation of $(2+1)$-TQFTs based on modular categories such as $(E_8)_1$.  For modular categories from Chern-Simons theories, our models gap out the $F\wedge F$-theories in the bulk, while the Chern-Simosn theories on the boundary survive.

\subsection{Mathematical Underpinning and Continuous Limit}\label{mathpinning}

The conceptual underpinning of the $2D$ Levin-Wen model is two mathematical theorems:  The quantum double $\mZ(\mC)$ of a unitary fusion category $\mC$ is always modular, and the Turaev-Viro TQFT based on $\mC$ is equivalent to the Reshetikhin-Turaev TQFT based on $\mZ(\mC)$ \cite{Mueger}\cite{TuVi}\cite{BenK}.  Therefore, Levin-Wen model is a Hamiltonian realization of both theorems simultaneously.  The mathematical theory behind the new models should be a generalization of these theorems to $(3+1)$-dimension.

There is a structure theorem for premodular categories.  An object $x$ in a premodular category is transparent by definition if $\tilde{s}_{xy}=d_x d_y$ for any object $y$, where $\tilde{s}_{xy}$ is the topological invariant of the Hopf link colored by $x$ and $y$.  Transparent elements of a premodular category $\mC$ form a symmetric fusion category $S_{\mC}$.  By a theorem of Deligne, every symmetric fusion category is equivalent to the representation category of a pair $(G,\mu)$, where $G$ is a finite group and $\mu$ is a central element of $G$ of order $\leq 2$ (see \cite{Ostrik}).  When $\mu$ is the identity, then there is a quotient $Q_{\mC}$ of $\mC$ by $S_{\mC}$, which is modular.  When $\mu$ is not the identity, a generalized quotient exists.  Therefore, essentially $\mC$ is some kind of extension of the quotient $Q_{\mC}$ by $S_{\mC}$.

The spin lattice models have continuous limits which are TQFTs based on pictures.  A framework for constructing picture TQFTs is formulated in \cite{Walker06}.  The ribbon graphs based on a premodular category are examples of a system of fields in the sense of \cite{Walker06} and hence lead to a $(3+1)$-TQFT.  In general, picture TQFTs have state-sum formulation, which can be realized by spin models.

\subsection{Holographic Resolution of Anomaly}

$(2+1)$-Witten-Chern-Simons (WCS) theories have anomaly in the sense that the path integral for closed 3-manifolds are not well-defined unless the $3$-manifolds are endowed with some extra structures such as $2$-framing \cite{Witten89}.  In dimension $2$ the anomaly is manifested in the chiral central charge of the boundary CFT.  The $(3+1)$-TQFT based on a modular category can be regarded as a holographic resolution of the anomaly: the $(2+1)$-WCS theories are really $(3+1)$-TQFTs.  Our model leads to a holomorphic tensor network representation for Chern-Simons theories, which is probably related to some generalization of the AdS/CFT correspondence.

Given a modular category $\mC$, the Reshetikhin-Turaev (RT) TQFT leads to a topological invariant of a 3-manifold $X$ with some extra-structure.  A convenient way to encode this extra structure is by an integer $n$.  Therefore, by an extended 3-manifold, we mean a pair $(X,n)$, where $X$ is an oriented $3$-manifold and $n$ an integer.  For a surface $Y$, the extra structure can be given a Lagrangian subspace $L$ in $H_1(Y;\Q)$.  RT TQFT will associate a vector space to the pair $(Y,L)$.  From the $(3+1)$-point of view, the extra structures $n$ and $L$ give instructions to finding manifolds one-dimensional higher.  Then the associated path integral and vector space for these one-dimension higher manifolds are exactly the path integral and vector space from RT TQFTs.  Specifically, given an extended $3$-manifold $(X,n)$, choose an oriented $4$-manifold $W$ bounding $X$ and $\sigma(W)=n$, then $Z_{3+1}(W)=Z_{RT}((X,n)).$  For an extended surface $(Y,L)$, choose a $3$-manifold $X$ so that the kernel of the inclusion of $H_1(Y;\Q)$ in $H_1(X;\Q)$ is $L$, then $Z_{3+1}(X)=Z_{RT}((Y,L))$.

\section{Applications}

\subsection{$3D$ Topological Insulators and BF theories}

There are several proposals for the effective TQFTs that model topological insulators \cite{QHS}\cite{CM}.  The modeling with BF theories is of particular interests to us because $(3+1)$-BF theories are related to $(3+1)$-TQFTs based on unitary modular categories.

Consider the $(3+1)$-BF theory with $G=SU(2)$.  By integrating out the $B$-field when $\Lambda\neq 0$, we have
$$Z(W)=\int \int_{B,A}e^{i \int_W Tr(B\wedge F+\frac{\Lambda}{12}F\wedge F)}DBDA=\frac{2 \pi }{\sqrt{\Lambda}}\int_A e^{-\frac{3i}{\Lambda}\int_W Tr(F\wedge F)}DA.$$
Comparing with the $SU(2)$-WCS theory, we see that $\frac{12 \pi}{\Lambda}$ corresponds to the level $k$ in $SU(2)$-WCS theories.  It is generally believed that the path integrals in the top dimension of a TQFT determine the extended TQFTs down to at least codimension$=2$.  If so, then $(3+1)$-BF theories are the same as WCS theories promoted to $(3+1)$.  When $\Lambda \rightarrow 0$, which is the same as $k\rightarrow \infty$, the limit theory is a semi-classical one.

As is explained in earlier sections, it is possible to compute statistics of the elementary extended excitations in the $(3+1)$-TQFTs, though in practice it is not an easy task.  Physically, we can couple the rank$=2$ tensor field in equ. (30) of \cite{CM} with extended objects such as a line field for $G=SU(2)$ (choose a $\Lambda$ corresponding to some level $k$, say $k=2$, in $SU(2)$-WCS theory.)  Conjecturally, the resulting statistics will be the same as computed from the mathematical theories.  It will be interesting to check some examples.

More interestingly, there are several proposals for potential fractional topological insulators \cite{MQKS} \cite{SBMS}.  As alluded in the introduction, we believe that the underlying topological orders in the fractional topological insulators when symmetry is broken will correspond to some $(3+1)$-TQFTs based on UBFCs.  It will be interesting to find out what are the corresponding UBFCs for the proposed fractional topological insulators.

\subsection{Projective Ribbon Permutation Statistics}

In \cite{FHNQWW}, the possibility of ribbon permutation statistics is studied for a collection of confined pointed excitations, called hedgehogs.  If the hedgehogs become deconfined in some related theory, then an effective description by a $(3+1)$-TQFT is a possibility.  We conjecture that one such possible deconfined phase could be a $(3+1)$-TQFT based on the following unitary braided fusion category.

Fermionic quantum Hall liquids can be described using spin modular categories \cite{RW}. For $\frac{5}{2}$-FQH liquids, the fermionic Moore-Read state with $6$-fold degeneracy on the torus is covered by a rank$=12$ UBFC.  This category is the even sector of  $Ising\times \Z_8$: the direct product of the Ising theory and a modular theory with fusion rules $\Z_8$.  The Ising theory has anyon types $\{1,\sigma, \psi\}$ and the anyon types of
the modular $\Z_8$ theory are denoted by $\{0,1,...,7\}$.  The anyon $f=\psi\otimes 4$ is a fermion.  The even sector, the anyons which are local with respect to the fermion $f$, consists of $\{\one\otimes i,\psi\otimes i\}$ for $i=$even and $\{\sigma \otimes i\}$ for $i=$odd.  This even sector is a rank$=12$ UBFC, therefore it leads to a $(3+1)$-TQFT.

\section{Further discussion}

The most general TQFTs are given by universal manifold pairing \cite{FKNSWW}.  This is tautology: the invariant of a manifold is itself considered as a vector in some vector space.  Dimension $4$ is again different from lower dimensions because the pairing has null (or light-like) vectors, while there are no  such null vectors in dimensions $1,2,3$.   If we restrict to unitary and anomaly-free TQFTs, then some smooth structures of some $4$-manifolds cannot be distinguished by the universal manifold pairing in dimension $4$.

We expect a true $3D$ generalization of the Levin-Wen model will be based on a generalization of unitary fusion category to a unitary fusion $3$-category and membranes.  Thus it might seem strange that interesting $(3+1)$-TQFTs can be constructed based on string-nets.  But the string-net that we are using are membranes: the strings are ribbons, so we are working with very simple membranes.  Still why the $6j$ symbols are useful algebraic input because the algebraic input should be solutions to a generalization of the pentagons to $3D$?  Braided tensor categories are examples of $3$-categories.  A curious fact is that the pentagons wear two hats: in $2D$, it is the algebraic equation for diagonal flip, while in $3D$, it is the Pachner $2$-$3$ move.  Presumably this is a manifestation of the relation between $2D$ CFTs and $3D$ TQFTs.

If no fermions are involved, it is not clear if a high category generalization of Levin-Wen model would produce $(3+1)$-TQFTs to distinguish smooth structures of $4$-manifolds.  Based on universal manifold pairing and Witten-Donadlson theory,
we would speculate that fermions are important for formulating $(3+1)$-TQFTs capable of distinguishing smooth structures.
In $2D$, a generalization of Levin-Wen model to include fermions is proposed in \cite{GWW}, which shows that topological orders in systems with fermions are strictly richer than purely bosonic systems.

The $N=2$ supersymemtric Yang-Mills theory for Witten-Donaldson theory is gapless \cite{WittenS}.  Moreover, both the $N=2$ and $N=4$ supersymemtric $(3+1)$-TQFTs are not unitary.  It seems to be an open question whether or not there are unitary and gapped $3+1$-TQFTs that can detect smooth structures of $4$-manifolds.

In an exposition for mathematicians \cite{Witten07}, Witten speculated about a possible connection between Seiberg-Witten theory and superconductors.  The challenge to better understand $(3+1)$-TQFTs lies at the frontier in both topology and the exploration of topological phases of matter.  A realization of Seiberg-Witten theory in condensed matter systems would be a landmark in the interaction of topology and physics.

\begin{appendix}

\end{appendix}

\end{document}